\documentclass[twocolumn, showpacs, amsmath,superscriptaddress, amsfonts,prb]{revtex4}
\usepackage{graphicx}
\begin{document}
\title{Field induced nucleation in the presence of a metal electrode}
\author{V. G. Karpov}
\email{victor.karpov@utoledo.edu}
\affiliation{Department of Physics and Astronomy, University of Toledo, Toledo OH 43606}
\author{R. E. E. Maltby}
\email{Robert.Maltby@rockets.utoledo.edu}
\affiliation{Department of Physics and Astronomy, University of Toledo, Toledo OH 43606}
\author{I. V. Karpov}
\email{ilya.v.karpov@intel.com}
\affiliation{Intel Corporation,2501 NW 229 Avenue, RA3-252, Hillsboro, OR, 97124}
\author{E. Yalon}
\email{eilamy@tx.technion.ac.il}
\affiliation{Electrical Engineering Department, Technion -
Israel Institute of Technology, Haifa 32000, Israel}
\begin{abstract}
    We consider the effect of metal electrodes on the polarizability and nucleation of metal phases responsible for the operations of the emerging solid state memory. Our analysis is based on the image charge approach. We find results for point dipoles in static and oscillatory fields as well as an erect cylindrical nucleus near metal surfaces in resistive switching memories. We predict a large increase in polarizability and nucleation rate due to the metal electrode effects.
\end{abstract}
\maketitle
\section{Introduction}

Operations of the phase change and filamentary resistive memory rely on the threshold switching when the electric bias induces the formation of conductive filaments. That phenomenon was explained as the electric field induced nucleation (FIN)\cite{karpov2007,karpov2008,karpov2008a,karpov2008b}
of needle shaped particles.  The large dipole moments $\mathbf{p}=\alpha\mathbf{E}$  of these particles provide an energy gain $F_E=-\mathbf{p}\cdot\mathbf{E}=\alpha E^2$ which outweighs the surface energy loss due to their non-spherical shape (here $\mathbf{E}$ is the external electric field and $\alpha$ is the polarizability). Following the first applications to the phase change memory, other applications of FIN later extended over resistive memory \cite{staikov2013,bernard2010,gonon2010,soni2011}, threshold switching, \cite{nardone2009} general metal particle nucleation, \cite{nardone2012} non-photochemical laser induced nucleation, \cite{nardone2012a} metal whisker nucleation, \cite{karpov2014} and resonance nucleation via plasmonic excitations of metal particles. \cite{karpov2012,grigorchuk2014}

In simple words, the FIN nucleation concept exploits the known fact \cite{EoCM} that the polarizability of a metal needle is proportional to the cube of its length, $\alpha\propto l^3$, instead of its volume $la^2$ where $a$ is the needle radius. As a result, starting from certain field strength, the electrostatic free energy gain $F_E\sim l^3E^2$ prevails over both the surface energy ($\propto la$) and the volume contribution ($\propto la^2$). The FIN predicted nucleation barrier is reciprocal in $E$ and the square root of the particle polarizability.

\begin{figure}[t!]\centering\includegraphics[width=0.55\linewidth]{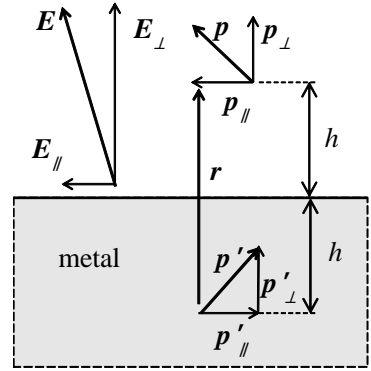}\caption{An arbitrarily oriented dipole in the presence of an arbitrary external field, near a metal surface.  The field ${\bf E}$ the dipole moment ${\bf p}$ and the image dipole ${\bf p'}$ are broken down into components parallel and perpendicular to the surface.}\label{Fig:dipole}
\end{figure}

The existing FIN theory assumes an unspecified strong external field ${\bf E}$ that remains unchanged in the course of nucleation. However the device metal electrodes will respond to the formation of a particle dipole ${\bf p}$ by generating the image dipole \cite{EoCM} ${\bf p'}$ as illustrated in Fig. \ref{Fig:dipole}. The interaction between these dipoles will contribute to the system polarization and thereby invalidate the concept of a fixed external field.

The image charge feedback will increase the bare dipole polarization thus lowering its electrostatic energy and accelerating FIN. A qualitatively similar phenomenon of the enhanced electromagnetic field was discussed earlier for Raman scattering by absorbed molecules. \cite{king1978,mohamed2009}

\begin{figure}[h!]\centering\includegraphics[width=0.4\textwidth]{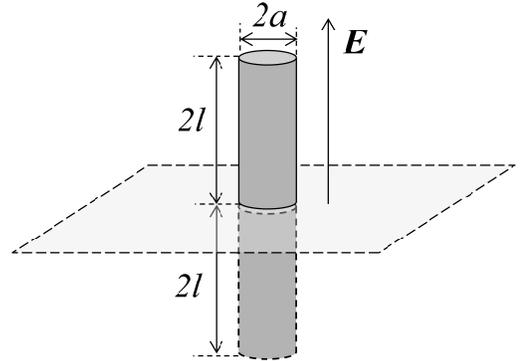}\caption{A right circular cylinder (height $2l$ and radius $a\ll l$) extending from a metal surface (the dashed parallelogram) in the presence of a constant electric field normal to the surface.  Also shown is the cylinder's image (dashed).}\label{Fig:cylinder}
    \end{figure}

Here, we present a self-consistent theory of FIN  with the image dipole taken into account. Our findings will represent a second example of image charge effects in device physics following the famous effect of image charge lowering of the Schottky barriers between semiconductors and metals. \cite{sze}

More specifically, in Sec. \ref{sec:interaction} we show that the interaction energy consists of that  between the bare field and the real dipole, ${\bf p\cdot E}$ and that of dipole-dipole, ($\propto {\bf p\cdot p'}$) interaction while that between the field and the image dipole does not contribute. Sec. \ref{sec:pointdip} presents the exactly solved case of point dipoles (far enough from the metal electrode), which shows that the image dipole effect can exponentially accelerate FIN. In Sec. \ref{sec:erect} we determine the electrode effect on the nucleation energy of an erect cylinder shaped particle nucleated perpendicular on the flat electrode as illustrated in Fig. \ref{Fig:cylinder}.
In that case, which is especially important for memory device operations, based on resistive switching, also known as RRAM and CBRAM, we show that the image charge effect can decrease the nucleation barrier by approximately a factor of 2, thus accelerating the FIN rate by many orders of magnitude
due to the exponential dependence of nucleation probability on the barrier.
Sec. \ref{sec:ac} analyzes the image charge effects for the case of ac fields which can be in resonance with the plasmonic excitation of nucleated particles, making their polarizability anomalously high. Finally, Sec. \ref{sec:concl} summarizes the results and discusses their limitations and possible complimentary effects.

\section{Interaction near a metal plate}\label{sec:interaction}

The superposition of the external field $\mathbf{E}$ (above the electrode), the field $\mathbf{E_r}$ due to real charges (e.g. the electric dipole), and the field $\mathbf{E_i}$ due to image charges, corresponds to the electrostatic energy $(1/8\pi )\int d^3r({\bf E}+{\bf E}_r+{\bf E}_i)^2$ where the integral is taken over the entire space outside the metal. Subtracting the partial energies, $(1/8\pi )\int d^3r[({\bf E})^2+({\bf E}_r)^2+({\bf E}_i)^2]$, yields the energy of interaction,
\begin{equation}\label{eq:inter}
W=\frac{1}{4\pi}\int d^3r({\bf E}\cdot {\bf E}_r+{\bf E}_i\cdot {\bf E}_r+{\bf E}\cdot {\bf E}_i).
\end{equation}

Taking into account the standard electrostatic equations, ${\bf E}=-\nabla \phi$ with $\phi =-{\bf E}\cdot {\bf r}$, and $\nabla ^2 \phi _{r,i}=-4\pi \rho _{r,i}$ for the potentials $\phi _{r,i}$ and charge densities $\rho _{r,i}$ of the real (r) and image (i) charges, one can write $W$ in the form
\begin{equation}\label{eq:inter1}
W=-{\bf E}\cdot {\bf p}-\int d^3r\phi _i({\bf r})\rho _r({\bf r}),
\end{equation}
where the real charge dipole moment is ${\bf p}=\int d^3r {\bf r}\rho _r({\bf r})$.

The third term in Eq. (\ref{eq:inter}) gives zero contribution to the result in Eq. (\ref{eq:inter1}) because the electric field is zero at the image charge density $\rho _i({\bf r})$ inside the metal. The interpretation of that zero contribution in terms of polarization {\it surface} charges (represented above with the image charges), is that they are redistributed along the equipotential metal surface.

We now specify the above Eq. (\ref{eq:inter1}) for the two cases analyzed in what follows. For the case of a spatially confined charge distribution, such as a point dipole, one can represent $\rho _r({\bf r})= -\nabla {\bf P}$ where ${\bf P}$ is the polarization density, and $\int {\bf P}d^3r={\bf p}$. Integrating by parts, this yields,
\begin{equation}\label{eq:Upd}
W=-({\bf E}+{\bf E}_{p'})\cdot {\bf p} \quad {\rm with} \quad {\bf E}_{p'}=\frac{3({\bf p'\cdot r}){\bf r}-{\bf p'}r^2}{r^5}
\end{equation}
where ${\bf E_{p'}}=-\nabla \phi _i$ is the field induced by the image dipole moment ${\bf p'}$.

For the case of an erect nucleus on the electrode (Fig. \ref{Fig:cylinder}), we note that the total electric potential along that metal rod must be a constant. That results in $\phi _i=-\phi _0={\bf E}\cdot {\bf r}$. Substituting this into Eq. (\ref{eq:inter1}) yields
\begin{equation}\label{eq:Uerect}
W=-2{\bf E}\cdot {\bf p}.
\end{equation}
This doubling of the well known result $U=-\mathbf{E}\cdot\mathbf{p}$ for a stand alone dipole can be explained physically as the attraction of the dipole to its image.

Eqs. (\ref{eq:Upd}) and (\ref{eq:Uerect}) are not final: the quantities, ${\bf p'}$ and ${\bf p}$ in them must be self-consistently determined, which is done in what follows. As long as $p$ is linear in $E$, this will define $U$ as a function quadratic in the bare field strength $E$ identified with the electrostatic contribution to free energy directly related to the particle nucleation barrier,
\begin{equation}\label{eq:effpol}
W=-\alpha _{\rm eff}E^2
\end{equation}
 where the effective polarizability $\alpha _{\rm eff}$ depends on the particle geometry.

Finally, we note that taking into account that the source dipole is in a material with dielectric permittivity $\varepsilon$ will not change the image charge values and interactions. However, it was shown by several authors \cite{kaschiev2000,warshavsky1999,isard1977} that the expression for the electrostatic free energy will change to
\begin{equation}\label{eq:FE}F_E=\varepsilon W.\end{equation}

\section{Point dipoles}\label{sec:pointdip}

The equations for the dipole components parallel ($p_{\parallel}$) and perpendicular ($p_{\perp}$) to the metal surface, take the form
\begin{equation}\label{eq:selfcon}
p_\parallel=\alpha _\parallel(E_\parallel +E_{p',\parallel}),\quad p_\perp=\alpha _\perp (E_\perp +E_{p',\perp}).
\end{equation}
Here  $\alpha _\parallel$ and $\alpha _\perp$ are respectively longitudinal and transverse polarizability. The corresponding components of the field generated by the image dipole at the real dipole location are given by
\begin{equation}\label{eq:dipoles}
E_{p',\parallel}=-\frac{p_\parallel '}{r^3}=\frac{p_\parallel}{r^3}\quad {\rm and}\quad E_{p',\perp}=\frac{2p_\perp '}{r^3}=\frac{2p_\perp}{r^3}
\end{equation}
where we have taken into account that vector ${\bf r}$ in Eq. (\ref{eq:Upd}) is perpendicular to the metal plane, and that $p_\perp '=p_\perp$, $p_\parallel '=-p_\parallel$. The latter relations illustrated in Fig. \ref{Fig:dipole} are readily justified by considering the image charges dual to  those forming the bare dipole.

Substituting Eq. (\ref{eq:dipoles}) into Eq. (\ref{eq:selfcon}) yields the expression for the effective (apparent) polarizabilities that determine the induced dipole moments,
\begin{equation}\label{eq:pE}
p_\parallel = (\alpha _\parallel )_{\rm eff} E_\parallel\quad {\rm and}\quad p_\perp = (\alpha _\perp )_{\rm eff} E_\perp
\end{equation}
with
\begin{equation}\label{eq:alphas}
(\alpha _\parallel )_{\rm eff}=\frac{\alpha _\parallel}{1-\alpha _\parallel /4h^3}\quad {\rm and}\quad  (\alpha _\perp )_{\rm eff}=\frac{\alpha _\perp}{1-\alpha _\perp /2h^3}.
\end{equation}
The latter equations are similar to the results of Ref. \onlinecite{king1978} describing the Raman scattering of molecules on metal surfaces.

Finally, substituting Eqs. (\ref{eq:pE}) and (\ref{eq:alphas}) into Eq. (\ref{eq:Upd}) gives the interaction energy
\begin{equation}\label{eq:Upd1}
W=-\frac{E_\parallel ^2\alpha _\parallel (1+3\alpha _\parallel /4h^3)}{(1-\alpha _\parallel /4h^3)^2}-\frac{E_\perp ^2\alpha _\perp (1+3\alpha _\perp /2h^3)}{(1-\alpha _\perp /2h^3)^2}.
\end{equation}
Eq. (\ref{eq:Upd1}) can be reduced to the form of Eq. (\ref{eq:effpol}) given the angle that the field ${\bf E}$ makes with the electrode. Overall, the results in Eqs. (\ref{eq:alphas}) and (\ref{eq:Upd1}) clearly show the metal induced enhancement in polarization and interaction energy compared to the case of stand alone particles corresponding to $h\rightarrow \infty$.

It was shown in the preceding work \cite{karpov2008,karpov2008a,karpov2008b} that FIN nucleation barrier is inversely proportional to the electric field strength. Due to the electrode, the latter is stronger by the image dipole contribution. The corresponding barrier suppression factor, i. e. the ratio of FIN barriers for stand alone particle over that in electrode proximity,  can be derived from Eqs. (\ref{eq:alphas}) and (\ref{eq:Upd1}) to be
\begin{equation}\label{eq:barriersuppr}
\sqrt{-\frac{E_\parallel ^2+E_\perp ^2}{W}}>1.
\end{equation}
Therefore, FIN is strongly accelerated in the closest vicinity of a metal electrode.

To estimate the latter suppression more quantitatively, we consider the case of needle shaped dipoles of length $l$ perpendicular to the metal plate, for which $\alpha _\perp\sim l^3$ to the accuracy of logarithmic corrections (see Refs. \onlinecite{karpov2007,karpov2008,karpov2008a,karpov2008b,nardone2009,nardone2012,nardone2012a,karpov2014,karpov2012,grigorchuk2014} and p. 17 in Ref. \onlinecite{EoCM}). Based on the general argument and the results in Refs. \onlinecite{karpov2007,karpov2008,karpov2008a,karpov2008b,nardone2009,nardone2012,nardone2012a,karpov2014,karpov2012,grigorchuk2014}, one can estimate the electrode induced nucleation barrier decrease as
\begin{equation}\label{eq:deltaW}
\delta W_B\sim W_B\frac{\alpha _{\rm eff}-\alpha _\perp}{\alpha _\perp}\sim W_B\left(\frac{l}{h}\right)^3
\end{equation}
where $W_B$ is the barrier for stand alone particles. This increases the nucleation rate by the factor $\exp(\delta W_B/kT)$ where $kT$ is the thermal energy.  Even though $l\ll h$ for point dipoles, the ratio $W_B/kT\lesssim 100$ can be so large that the acceleration factor $\exp(\delta W_B/kT)$ amounts to several orders of magnitude.

\section{Erect cylinder}\label{sec:erect}

It follows from the previous section that nucleation of a conducting cylinder erect on a metal electrode is a most likely scenario to conductive path creation in an emerging solid state memory. We now specify the result in Eq. (\ref{eq:Uerect}) describing the energy of such a nucleus. First of all, we note that the dipole moment ${\bf p}$ in that equation corresponds to the half of the total cylinder comprising both the real and image dipoles,
\begin{equation}\label{eq:pptot}
{\bf p}=\frac{{\bf p}_{\rm tot}}{2}.
\end{equation}

On the other hand, the electric field distribution in the system is such as it would be in the case of the latter stand alone total cylinder, as follows from the system symmetry (see p. 45 of Ref. \onlinecite{batygin1978}). Following Ref. \onlinecite{EoCM} (p. 17), the dipole moment of such a cylinder of length $4l$ (see Fig. \ref{Fig:cylinder}) can be written as
\begin{equation}\label{eq:pl}
{\bf p}_{\rm tot}=\frac{(4l)^3{\bf E}}{3\ln(32l/a)+7}.
\end{equation}
Combining this with Eq. (\ref{eq:pptot}) yields
\begin{equation}\label{eq:pfinal}
p=4p_0\frac{3\ln(8l/a)+7}{3\ln(32l/a)+7}\approx 4p_0
\end{equation}
where $p_0$ is the dipole moment of a stand alone cylinder of length $2l$ infinitely far from the electrode and we have taken into account that both the numerator and denominator of the ratio in Eq. (\ref{eq:pfinal}) are supposed to be much greater than unity. \cite{EoCM} In other words, the polarizability of a perpendicular cylinder erect on the electrode is four times of that for a stand alone cylinder of the same geometry.

Following the existing FIN theory, the nucleation barrier is reciprocal in the square root of polarizability [see e. g. Eq. (10) in Ref. \onlinecite{karpov2008}]. Therefore, the presence of a metal electrode suppresses the nucleation barrier by a factor of 2. Because the nucleation rate is proportional to $\exp(-W_B/kT)\ll 1$, such a suppression will accelerate FIN by many orders of magnitude. This enables one to make more accurate numerical comparison to the data and conclude that the field induced nucleation of conductive phases is exponentially more efficient in the vicinity of a metal electrode.

 \section{Point dipoles in a resonance field}\label{sec:ac}

While consideration in this section is not directly related to device operations, it is closely linked to the point dipole polarization effects that may appear diverging when the electric field is in resonance with their internal excitations. Indeed, Eq. (\ref{eq:alphas}) is limited to the concept of point dipoles in that it assumes their small geometrical size, but not necessarily a small polarizability. As follows from the general argument and was explicitly shown in multiple work on  plasmonic effects (see e. g. Refs. \onlinecite{karpov2012,maier2007} and references therein), internal excitations in resonance with an ac field can significantly increase $\alpha$ compared to its static value $\sim l^3$, to the extent that the denominators in Eq. (\ref{eq:alphas}) nullify making $(\alpha _\perp)_{\rm eff}$ and $(\alpha _\parallel)_{\rm eff}$ divergent.

For specificity, we describe here how this problem is solved for the case of dipoles parallel to the metal surface. Taking into account Eq. (\ref{eq:alphas}) and following the known recipe for the energy of a metal particle in an ac field, \cite{EoCM,bohren1983} the contribution to the free energy of a metal particle becomes
\begin{equation}\label{eq:acenergy}
F_E=\frac{E_\parallel ^2}{2}{\rm Re}\left(\frac{\alpha _\parallel}{1-\alpha _\parallel /4h^3}\right),\quad \alpha _\parallel =\frac{V}{4\pi}\frac{\varepsilon _p-\varepsilon}{\varepsilon +n(\varepsilon _p-\varepsilon)}.
\end{equation}
Here, $\epsilon _p=1-\omega _p^2/\omega ^2+i\omega _p^2/\omega ^3\tau$
is the dielectric permittivity of a metal particle of volume $V$ at frequency $\omega$, $\tau$ is the electron relaxation time, $\omega _p$ is the plasmon frequency, $n$ is the depolarizing factor, and $\tau ^{-1}\ll\omega\ll\omega _p$. In Eq. (\ref{eq:acenergy}) $E_\parallel$ is understood as the amplitude of the field component parallel to the metal surface. Approximating the dipole geometry with anisotropic prolate spheroid of semi-axes $l$ and $a\ll l$ yields,
\begin{equation}\label{eq:depfac}n\approx (a/l)^2[\ln(2l/a)-1] \quad {\rm and}\quad  V=4\pi la^2/3.\end{equation}

For the case of stand alone ($h\rightarrow \infty$) particles, the algorithm of Eq. (\ref{eq:acenergy}) has been implemented. \cite{karpov2012,grigorchuk2014} Along exactly the same lines, we have analyzed here the case of finite $h$, which is still straightforward, though more cumbersome.  Omitting the derivation, which is quite similar to that in Refs. \onlinecite{karpov2012,grigorchuk2014}, the final result is rather transparent: compared to the case of stand alone particle, $F_E$ turns out to be renormalized by the coefficient  where $\alpha _\parallel$ is the {\it static} ($\omega =0$) polarizability; hence, there is no apparent resonance divergency.

In the meantime, $F(E)$ remains to be enhanced compared to the static field case due to the resonance interaction with plasmonic excitations in metal nuclei, exactly as established earlier \cite{karpov2012,grigorchuk2014} for the stand alone needle shaped particles. It is additionally enhanced by the image dipole related factor $1/(1-\alpha _\parallel /4h^3)$. Similar conclusions (with obvious modifications) can be made for the dipoles perpendicular to the metal surface.

\section{Discussion and Conclusions}\label{sec:concl}

\begin{figure}[thb]\centering\includegraphics[width=1.1\linewidth]{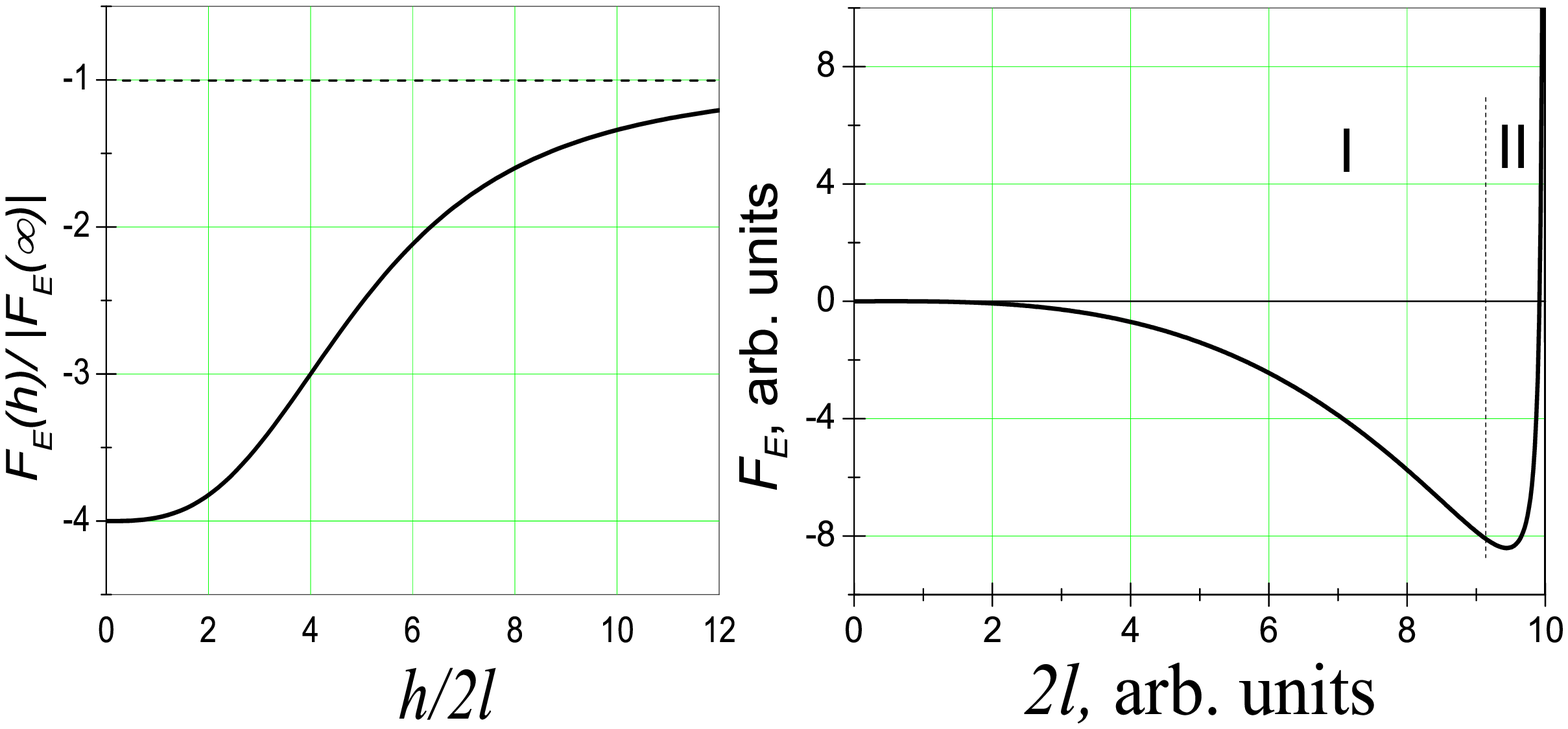}\caption{{\it Left}: The electrostatic energy for a perpendicular metal cylinder of length $2l$ at distance $h$ from the electrode; in units of the absolute value of energy of a stand alone cylinder $|Ep_0|$. {\it Right}: The electrostatic energy vs. length $2l$ of a perpendicular metal cylinder erect on an electrode in a system where a second parallel electrode is placed at distance $20l$.}\label{Fig:summary}
\end{figure}

Fig. \ref{Fig:summary} represents a summary of our results, along with an additional feature for the case of a conductive filament reaching almost entirely through the device thickness.  The left diagram in Fig. \ref{Fig:summary} shows how the dipole energy follows Eq. (\ref{eq:Upd1}) for the case when the distance to the electrode is much greater than the dipole geometrical length, but for the case of smaller distances, comparable to the dipole's length, the dipole energy saturates and becomes equal to the energy of a perpendicular cylinder erect on the electrode, $-4Ep_0$ when $h=0$.

The right diagram in Fig. \ref{Fig:summary} shows how the electrostatic energy of a cylinder erect on the device electrode follows the dependence $F_E=-pE\propto -l^3$ with $p$ given by Eqs. (\ref{eq:pptot}) and (\ref{eq:pl}) until the cylinder length becomes very close to the device thickness $L$ (taken to be $20l$ in the diagram). This implies that the presence of the second electrode remains immaterial (and the above results apply in domain I of the diagram) until the cylinder almost bridges the system. In the latter narrow complimentary domain II, the presence of the second electrode becomes important and can qualitatively change the electrostatic energy behavior from decreasing to sharply increasing as explained next.

The reason for a relatively small effect of the second electrode in domain I is that the tip of a needle shaped particle significantly disturbs the field only in a relatively small region of linear size \cite{karpov2014a} $\delta\sim ln$ where, according to Eq. (\ref{eq:depfac}), the depolarizing factor $n\ll 1$. The corresponding field concentration, \cite{EoCM,karpov2014a} $E\rightarrow E/n$ results in a rather insignificant energy contribution $\sim (E/n)^2\delta ^3\sim E^2l^3n\ll E^2l^3$. It can be neglected unless the filament length $l$ becomes very closed to $L$, separated by distance $\delta \ll L$ to the second electrode. Assuming the aspect ratios $l\gtrsim 10$ typical of the solid state memory filaments, $\delta$ is less than 1\% of the device thickness. Note that $\delta$ is not related to the width of domain II in Fig. \ref{Fig:summary}.

To explain the origin of domain II, we note that the above reasoning is limited to the assumption of electrostatic conditions where the currents are absent and the charges remain fixed. Correspondingly, the electrostatic energy is described by $Q^2/2C$ where $Q$ is the electric charge and $C$ is the capacitance. With $C$ increasing upon nucleation of metal particles, the electrostatic energy decreases, leading to the free energy gain described above.

These electrostatic conditions will be violated when the distance $d\ll l$ between the particle tip and the second electrode becomes small enough to allow efficient current flow by tunneling or hopping. In that case, the electric potential difference $U$ between the tip and the electrode rather then the field or electric charge becomes essential. Correspondingly, the electrostatic energy is better represented as $CU^2/2$ where $C\sim a^2/d$ is the effective capacitance between the tip and the electrode. The right diagram in Fig. \ref{Fig:summary} shows the divergent contribution $a^2U/d$ in domain II where the results of the preceding sections do not apply.

While the latter analysis remains rather qualitative, its predicted energy barrier in domain II can have significant implications limiting the filament length and enforcing a gap between the filament tip  and the electrode. That prediction is consistent with the published experimental data on resistive memory establishing the existence of the gap. \cite{yalon2012,yalon2012a}

Our general conclusion is that the image charge effects greatly facilitate nucleation of metal particles leading to the acceleration of the FIN rates by many orders of magnitude in the near electrode region. These effects are generic, i. e. they do not depend on the material chemistry and apply to all conductive electrodes.

From the practical standpoint, the 50\% reduction in the FIN barrier by image charge effects explains why the conductive filaments always nucleate at the electrode rather than in the bulk of the device, and why FIN can occur even for the case of electrodes with relatively low roughness.

In addition, our results show how the resonance FIN under the laser beam is exponentially more likely in the near electrode region; this should be taken into account in planning experiments on non-photochemical laser induced nucleation. \cite{garetz2002,ward2012}

We are grateful to E. Yalon and A. B. Pevtsov for useful discussions.


\begin{thebibliography}{99}
\bibitem{karpov2007}V. G. Karpov, Y. A. Kryukov, S. D. Savransky, and I. V. Karpov, Nucleation Switching in Phase Change Memory, Appl. Phys. Lett. {\bf 90}, 123504 (2007).
\bibitem{karpov2008}V. G. Karpov, Y. A. Kryukov, I. V. Karpov, and M. Mitra, Field induced nucleation in glasses, Phys. Rev. B {\bf 78}, 052201 (2008).
\bibitem{karpov2008a}I. V. Karpov, M. Mitra, D. Kau, G. Spadini, Y. A. Kryukov, and V. G. Karpov, Evidence of field induced nucleation in phase change memory, Appl. Phys. Lett. {\bf 92}, 173501 (2008).
\bibitem{karpov2008b}V. G. Karpov, Y. A. Kryukov, I. V. Karpov, and M. Mitra, Crystal nucleation in phase change memory, J. Appl. Phys. {\bf 104}, 054507 (2008).
\bibitem{staikov2013} I. Valov and G. Staikov, Nucleation and growth phenomena in nanosized electrochemical systems for resistive switching memories, J Solid State Electrochem {\bf 17}, 365 (2013).
\bibitem{bernard2010}Y. Bernard, P. Gonon, and V. Jousseaume, Resistance switching of Cu/SiO$_2$ memory cells studied under voltage
and current-driven modes, Appl. Phys. Lett., {\bf 96}, 193502 (2010).
\bibitem{gonon2010}P. Gonon, M. Mougenot, C. Vallée, C. Jorel, V. Jousseaume, H. Grampeix, and F. El Kamel, Resistance switching in HfO$_2$ metal-insulator-metal devices, J. Appl. Phys. {\bf 107}, 074507 (2010).
\bibitem{soni2011}R. Soni, P.  Meuffels, G. Staikov, R. Weng, C. Kügeler, A. Petraru, M. Hambe, R. Waser, H. Kohlstedt, On the stochastic nature of resistive switching in Cu doped Ge0.3Se0.7 based memory devices, J. Appl. Phys., {\bf 110}, 054509 (2011).
\bibitem{nardone2009}M. Nardone, V. G. Karpov, C. Jackson, and I. V. Karpov, Unified Model of Nucleation Switching, Appl. Phys. Lett. {\bf 94}, 103509 (2009)
\bibitem{nardone2012}M. Nardone and V. G. Karpov, Nucleation of Metals by Strong Electric Fields, Appl. Phys. Lett. {\bf 100}, 151912 (2012).
\bibitem{nardone2012a}M. Nardone and V. G. Karpov, Phenomenological theory of non-photochemical laser induced nucleation, Phys. Chem. Chem. Phys. {\bf 14}, 13601 (2012).
\bibitem{karpov2014} V. G. Karpov, Electrostatic theory of metal whsikers, Phys. Rev. Applied, {\bf 1}, 044001  (2014).
\bibitem{karpov2012}V. G. Karpov, M. Nardone, and N. I. Grigorchuk, Plasmonic Mediated Nucleation of Nanoparticles, Phys. Rev. B {\bf 86}, 075463 (2012).
\bibitem{grigorchuk2014}N. I. Grigorchuk and V. G. Karpov, Light induced nucleation of metallic nanoparticles with frequency controlled shapes, Appl. Phys. Lett., {\bf 105}, 223103 (2014).
\bibitem{EoCM} L. D. Landau and E. M. Lifshitz, Electrodynamics of Continuous Media (Pergamon, Oxford, New York, 1984).
\bibitem{king1978}F. W. King, R. P. Van Duyne, and G. C. Schatz, Theory of Raman scattering by molecules adsorbed on electrode surfaces, J. Chem. Phys., {\bf 69}, 4472 (1978).
\bibitem{mohamed2009}M. A. Mohamed, E. F. Kuester, and M. Piket-May, C. L. Holloway, The field of an electric dipole and the polarizability of a conducting object embedded in the interface between dielectric materials, Progress In Electromagnetics Research B, {\bf 16}, 1, (2009).
\bibitem{sze} S. M. Sze, {\it Physics of Semiconductor Devices} (Willey \& Sons, New York, 1981).
\bibitem{kaschiev2000}D. Kaschiev, {\it Nucleation: Basic Theory with Applications} (Butterworth-Heinemann, Oxford, Amsterdam, 2000).
\bibitem{warshavsky1999}V.B. Warshavsky, A.K. Shchekin, The effects of the external electric feld in thermodynamics of
formation of dielectric droplet., Colloids and Surfaces A: Physicochemical and Engineering Aspects {\bf 148}, 283 (1999).
\bibitem{isard1977}J.O. Isard, Calculation of the influence of an electric field on the free energy of formation of a nucleus, Phil. Mag. {\bf 35}, 817 (1977).
\bibitem{batygin1978}V. V. Batygin and I. N. Toptygin, {\it Problems in Electrodynamics}, Second Edition, Academic Press, London (1978).
\bibitem{maier2007}S. A. Maier, {\it Plasmonics: Fundamentals and Applications} (Springer, New York, 2007).
\bibitem{bohren1983}C. F. Bohren and D. R. Huffman, {\it Absorption and Scattering of Light by Small Particles}, Wiley, New York 1983.
\bibitem{karpov2014a}V. G. Karpov and Diana Shvydka, Semi-shunt field emission in electronic devices, Appl. Phys. Lett. {\bf 105}, 053904 (2014).
\bibitem{yalon2012}E. Yalon, A. Gavrilov, S. Cohen, D. Mistele, B. Meyler, J. Salzman, and D. Ritter, Resistive Switching in HfO2 Probed by a
Metal–Insulator–Semiconductor Bipolar Transistor, IEEE Electron. Device Lett., {\bf 33}, 11 (2012).
\bibitem{yalon2012a}E. Yalon, S. Cohen, A. Gavrilov and D. Ritter, Evaluation of the local temperature of conductive filaments in resistive switching
materials, Nanotechnology, {\bf 23}, 465201 (2012).
\bibitem{garetz2002} B. A. Garetz, J. Matic, and A. S. Myerson,  Polarization switching of crystal structure in nonphotochemical light-induced nucleation of supersaturated aqueous glycine solutions, Phys. Rev. Lett. {\bf 89}, 175501 (2002).
\bibitem{ward2012}M. R. Ward, S. McHugh and A. J. Alexander, Non-photochemical laser-induced nucleation of supercooled glacial acetic acid, Phys. Chem. Chem. Phys. {\bf 14}, 90 (2012).
\end{thebibliography}
\end{document}